\documentclass[prl,aps,a4paper,twocolumn,superscriptaddress,nofootinbib]{revtex4-2}
\usepackage[utf8x]{inputenc}
\usepackage{amssymb}
\usepackage{amsmath}
\usepackage{graphicx}
\usepackage{bbm}
\usepackage{psfrag}
\usepackage{latexsym}
\usepackage{hyperref}
\usepackage{color}
\usepackage{physics}
\usepackage{changepage}
\usepackage{scalerel}
\usepackage{verbatim}

\newcommand{\be}{\begin{equation}}
\newcommand{\ee}{\end{equation}}
\newcommand{\bea}{\begin{eqnarray}}
\newcommand{\eea}{\end{eqnarray}}
\newcommand{\ba}{\begin{eqnarray}}
\newcommand{\ea}{\end{eqnarray}}

\newcommand{\beq}{\begin{equation}}
\newcommand{\eeq}{\end{equation}}
\newcommand{\beqa}{\begin{eqnarray}}
\newcommand{\eeqa}{\end{eqnarray}}
\newcommand{\beqar}{\begin{eqnarray*}}
\newcommand{\eeqar}{\end{eqnarray*}}
\newcommand{\mc}[1]{\mathcal{#1}}
\newcommand{\bmc}[1]{\bar{\mathcal{#1}}}
\newcommand{\mfsl}{\mathfrak{sl}(2,\mathbb R)}
\newcommand{\mfsu}{\mathfrak{su}(2)}
\newcommand{\hsf}{\mathsf{h}}
\newcommand{\ssf}{\mathsf{s}}
\newcommand{\mTr}{\text{Tr}}
\newcommand{\msf}{\mathsf{m}}
\newcommand{\scalar}{\scaleto{\text{scalar}}{3.5pt}}

\newcommand{\gev}[1]{\big\langle #1\big\rangle_{ \scaleto{\text{grav}}{3.5pt}}}
\newcommand{\Li}[1]{\text{Li}_{#1}}

\newcommand{\ms}[1]{\mathsf{#1}}

\usepackage[normalem]{ulem}
\usepackage{soul}

\usepackage{etoolbox}
\preto\subequations{\ifhmode\unskip\fi}

\definecolor{purple}{rgb}{0.5,0.15,0.5}

\makeatletter
\newcommand*{\textlabel}[2]{%
  \edef\@currentlabel{#1}
  \phantomsection
  #1\label{#2}
}
\makeatother

\begin{document}

\author{Alejandra Castro}
\email{ac2553@cam.ac.uk}
\affiliation{\emph{Department of Applied Mathematics and Theoretical Physics, University of Cambridge,
Cambridge CB3 0WA, United Kingdom}}

\author{Ioana Coman}
\email{ioana.coman@ipmu.jp}
\affiliation{\emph{Kavli Institute for the Physics and Mathematics of the Universe (WPI), University of Tokyo,
Kashiwa, Chiba 277-8583, Japan}}

\author{Jackson R. Fliss}
\email{jf768@cam.ac.uk}
\affiliation{\emph{Department of Applied Mathematics and Theoretical Physics, University of Cambridge,
Cambridge CB3 0WA, United Kingdom}}

\author{Claire Zukowski}
\email{czukowsk@d.umn.edu}
\affiliation{\emph{Department of Physics and Astronomy, University of Minnesota Duluth,
Duluth, MN 55812, USA}}

\title{Coupling Fields to 3D Quantum Gravity via Chern-Simons Theory}
\begin{abstract}

We propose a mechanism that couples matter fields to three-dimensional quantum gravity, which can be used for theories with a positive or negative cosmological constant. 
Our proposal is rooted in the Chern-Simons formulation of three-dimensional gravity and makes use of the Wilson spool, a collection of Wilson loops winding around closed paths of the background. 
We show that the Wilson spool correctly reproduces the one-loop determinant of a free massive scalar field on rotating black holes in AdS$_3$ and Euclidean dS$_3$ as $G_N\to 0$.
Moreover, we describe how to incorporate quantum metric fluctuations into this formalism.

\end{abstract}

\maketitle

\section{Introduction}\label{sec:intro}

Chern-Simons theory is a compelling approach to quantum gravity that makes manifest the topological nature of gravity in three-dimensions \cite{Achucarro:1987vz, Witten:1988hc}.
For negative cosmological constant, $\Lambda<0$, Chern-Simons theory nicely characterizes the physics of black holes in Anti-de Sitter (AdS$_3$) space and their higher-spin generalizations \cite{Gutperle:2011kf,Ammon:2012wc,deBoer:2013gz}, and dovetails the AdS$_3$/CFT$_2$ dictionary with the existence of edge modes \cite{Coussaert:1995zp,Campoleoni:2010zq,Cotler:2018zff}. Importantly, however, Chern-Simons gravity provides a powerful computational foothold for three-dimensional quantum gravity without recourse to holography.  This is especially key in the context of $\Lambda>0$, de Sitter (dS$_3$) gravity, where the corresponding dS/CFT dictionary is much less understood; see, though, \cite{Cotler:2019nbi,Hikida:2021ese} for recent developments. This context exhibits the true efficacy of Chern-Simons gravity, allowing a characterization of loop corrections that are relevant to quantum cosmology \cite{Carlip:1992wg,Guadagnini:1995wv,Banados:1998tb, Park:1998yw, Govindarajan:2002ry,Castro:2011xb,Anninos:2020hfj,Anninos:2021ihe}.

It has been a long-standing problem to incorporate matter into Chern-Simons gravity while retaining the topological features that make it natural as a theory of quantum gravity.  In this letter we address this problem. In short, we introduce a new object we deem the {\it Wilson spool} which provides an effective coupling of massive fields to quantum gravity directly as a gauge-invariant operator. We define it precisely below, but intuitively, the spool represents a Wilson loop winding arbitrarily many times around a closed path for which the fields have non-trivial holonomy. This represents a pivotal entry into a dictionary mapping geometric quantities to quantum operators in Chern-Simons gravity  \cite{Witten:1989sx,Carlip:1989nz,Ammon:2013hba,Fitzpatrick:2016mtp,Castro:2018srf,Castro:2020smu}.  The Wilson spool will dictate how quantum gravity alters the physics of quantum fields in a manner that is quantitatively controlled by the gravitational coupling, $G_N$.

We will show that the Wilson spool is a natural object regardless of the sign of cosmological constant.  In the context of AdS$_3$ we will exactly reproduce, at tree-level $(G_N\to0)$, the one-loop determinant of a scalar field on a rotating BTZ black hole background. This computation is done directly ``in the bulk," without reference to holography. In fact, it is in the context of dS$_3$, where holography is of limited utility, that we can make use of the full power of this proposal. 
Certain exact results in Chern-Simons theory can be meaningfully adapted to accommodate features necessary for de Sitter gravity and its massive single-particle states.
This provides a principled and controlled method to computing $G_N$ corrections to one-loop determinants of fields coupled to dynamical gravity. In this letter, we distill key results that are presented in full detail in the companion paper~\cite{Castro:2023dxp},  and also cast them in a presentation that is unified for both signs of the cosmological constant.

In the following we will present our proposal of the Wilson spool, and illustrate its efficacy for AdS$_3$ and dS$_3$ gravity. We will evaluate it on AdS$_3$ in Lorentzian signature, where the Chern-Simons gauge group is $SL(2,\mathbb R)\times SL(2,\mathbb R)$, while  for dS$_3$ it will be treated in Euclidean signature, where the gauge group is $SU(2)\times SU(2)$. This choice facilitates making parallels between them, and  hence makes  evident the robustness of our proposal, while also showing contrast when appropriate.

\section{The proposal}\label{sec:spool}

 To illustrate our proposal it is instructive to cast some portions in the metric formulation.
Consider the path integral for a quantum scalar field $\phi$, of mass $\msf$,  with no self-interactions, and minimally coupled to the metric field, $g_{\mu\nu}$. This would be
\beq\label{eq:Zscalar1}
Z_{\scalar}[g_{\mu\nu}]=\int [\mc D\phi]e^{iS_{\rm matter}[\phi,g_{\mu\nu}]} ~.
\eeq
Including the perturbative, quantum fluctuations of the metric leads to
\beq\label{eq:Zmatter}
\gev{Z_{\scalar}[M]}:=\int[\mc Dg_{\mu\nu}]_{\scaleto{M}{3.5pt}}e^{-I_{\rm EH}[g_{\mu\nu}]}Z_{\scalar}[g_{\mu\nu}]~.
\eeq
For concreteness we wrote here a Euclidean path integral, where $I_{\rm EH}$ is the Euclidean Einstein-Hilbert action plus a cosmological constant term. The gravitational path integral is taken on a fixed topology $M$, and it includes all perturbative corrections around that topology. 

Our proposal for quantifying the coupling of matter to gravity is captured by the following equality,\footnote{Since both theories, AdS$_3$ and dS$_3$ gravity, involve the product of two gauge groups, we will introduce two Chern-Simons connections, $(A_L,A_R),$ corresponding to each group. These will encode the metric field $g_{\mu\nu}$ as we will review in the subsequent sections.}
\beq
Z_{\scalar}[g_{\mu\nu}]=\exp\left(\frac{1}{4}\mathbb W_j[A_L,A_R]\right)~.
\eeq
On the right hand side we have the Wilson spool
\begin{align}\label{eq:spooldef}
&\mathbb W_j[A_L,A_R]:=i\int_{\mc C}\frac{\dd\alpha}{\alpha}\frac{\cos\alpha/2}{\sin\alpha/2}\nonumber\\
&\qquad\qquad\times \mTr_{R_j}\left(\mc P e^{\frac{\alpha}{2\pi}\oint A_L}\right)\mTr_{R_j}\left(\mc Pe^{-\frac{\alpha}{2\pi}\oint A_R}\right)~,
\end{align}
which captures $Z_{\scalar}$ by using solely objects in Chern-Simons theory.
 Starting from the right, we have path ordered exponentials of $A_L$ and $A_R$ which encode $g_{\mu\nu}$ above: this is the portion that captures the information of the geometry. Next, we have traces over a representation $R_j$:  here is where we encode the single-particle representations of the field. The mass of the field is related to the Casimir of the representation
 \beq
 c_2=-\frac{\msf^2}{4\Lambda}~,
 \eeq
 with $\Lambda$ the cosmological constant. 
Finally, in \eqref{eq:spooldef} we have an integral over $\alpha$. The measure and contour ${\mc C}$ of this integral  serve two purposes \cite{Castro:2023dxp}: First, they make $\mathbb{W}$ free of UV divergences. Second, as will become clear in detail, evaluating this contour integral as a sum over poles implements a sum over Wilson loops with arbitrary winding, which is what makes it a ``spool.'' The specific details of ${\mc C}$, and allowed deformations, depend on the holonomies of $A_{L,R}$; these will be specified for the backgrounds considered in the following sections.

 In this letter we want to uphold \eqref{eq:spooldef} on two fronts. First, we believe \eqref{eq:spooldef} applies to any smooth three-dimensional background, including but also extending beyond dS$_3$ quantum gravity. Second, it is a useful expression to quantify quantum gravity effects: the Chern-Simons formulation allows us to integrate out the scalar field, and obtain an explicit functional in terms of the connections $A_{L,R}$. In this context, the main appeal of our proposal is the ability to quantify
\be\label{eq:Wgrav}
\begin{aligned}
& \gev{\log Z_{\rm scalar}}=\frac{1}{4}\gev{\mathbb W_j} \\
 &=\frac{1}{4}\int \mc D A_{L/R} \,e^{ik_L S_{\rm CS}[A_L]+ik_R S_{\rm CS}[A_R]} \mathbb W_j[A_L,A_R]~,
\end{aligned}
\ee
 where $S_{\rm CS}[A]$ is the Chern-Simons action, $k_{L/R}$ are the levels. For brevity, here $\mc D A_{L/R}$ accounts for the measure for the two copies of the gauge group.  The brackets mean that one is accounting for gravitational fluctuations around a fixed topology.  In the Chern-Simons language, this means that we fix a background holonomy for $A_{L,R}$ in addition to the topology.  It is important to stress that \eqref{eq:Wgrav} is a non-trivial function of $G_N$ and the mass of scalar field.

\section{Wilson spools in \texorpdfstring{A\lowercase{d}S$_3$}{AdS3} gravity}\label{sec:ads}

As a first example of the utility of our proposal we will focus on Chern-Simons gravity with negative cosmological constant, $\Lambda=-\ell_\text{AdS}^{-2}$. This is quantum gravity on spaces that are locally asymptotically Anti-de Sitter (AdS).  In Lorentzian signature, the relevant isometry group is $SL(2,\mathbb R)\times SL(2,\mathbb R)$ and massive particles in this space can be organized into $\mfsl$  representation theory.  The Lorentzian Einstein-Hilbert action is given by the difference in $SL(2,\mathbb R)$ Chern-Simons theories
\beq
S_\text{EH}=k\left(S_\text{CS}[A_L]-S_\text{CS}[A_R]\right)~,
\eeq
where $k=\frac{\ell_\text{AdS}}{4G_N}$.  The connections are related to the co-frame and the spin connection via
\beq\label{eq:ALAR-AdS}
A_L=(\omega^a+e^a/\ell_\text{AdS})L_a~,~~ A_R=(\omega^a-e^a/\ell_\text{AdS})\bar L_a~.
\eeq
Above $L_a$ and $\bar L_a$ generate the independent $\mfsl$ algebras.  Our conventions follow those in \cite{Gutperle:2011kf,Castro:2018srf}. 
At the classical level, the background geometries of interest are generated by flat background connections 
\begin{align}\label{eq:BTZbgconns}
a_L=&L_0\,\dd\rho+\left(e^{\rho}L_+-e^{-\rho}\frac{2\pi\mc L}{k}L_-\right)\dd x^+~,\nonumber\\
a_R=&-\bar L_0\,\dd\rho-\left(e^\rho\,\bar L_--e^{-\rho}\frac{2\pi\bmc L}{k}\,\bar L_+\right)\dd x^-~,
\end{align}
where $x^\pm=t\pm \varphi$ and $\varphi\sim\varphi+2\pi$. Upon using \eqref{eq:ALAR-AdS}, these are rotating BTZ black hole geometries \cite{Banados:1992wn,Banados:1992gq} with mass, $M$, and angular momentum, $J$, given by
\beq
\mc L=\frac{M\ell_\text{AdS}+J}{4\pi}~,\qquad\bmc L=\frac{M\ell_\text{AdS}-J}{4\pi}~.
\eeq
In Euclidean signature, we can rotate $(x^+,x^-)$ to complex coordinates $(z,-\bar z)$. Periodicity in $\varphi$ plus smoothness of the horizon implies these parameterize a complex torus, $(z,\bar z)\sim (z+2\pi m+2\pi n\tau,\bar z+2\pi m+2\pi n\bar\tau)$, $m,n\in\mathbb Z$, with modular parameter
\beq\label{eq:modpardef}
\tau =\frac{i}{2}\sqrt{\frac{k}{2\pi\mc L}}~,\qquad \bar\tau=-\frac{i}{2}\sqrt{\frac{k}{2\pi\bmc L}}~.
\eeq
What separates this complex torus from the torus defining the boundary of thermal AdS$_3$ is how we ``fill it in" in the bulk: in particular, the black hole geometry is filled in so that the thermal-cycle is bulk contractible while the spatial-cycle is not.  The holonomies of the background connections $a_{L/R}$ around this cycle, $\gamma_\varphi$, are easily computed to be
\begin{align}\label{eq:BTZholo}
\mc P\exp\left(\oint_{\gamma_\varphi}a_{L/R}\right)=&u_{L/R}^{-1}\,e^{i2\pi \hsf_{L/R} L_0}\,u_{L/R}~,
\end{align}
where $u_{L/R}$ are periodic group elements and
\beq\label{eq:BTZholo2}
\hsf_L=-\frac{1}{\tau}~,\qquad \hsf_R=-\frac{1}{\bar\tau}~.
\eeq

Next we will show that the Wilson spool of these background connections, $a_{L/R}$, reproduces the one-loop determinant of a massive scalar field on the BTZ background. That is, we will test the relation
\beq\label{eq:lZBTZtospool}
\begin{aligned}
\log Z_{\scalar}[\text{BTZ}]&=\log \det(-\nabla_{\scaleto{\rm BTZ}{3.5pt}}^2+\msf^2\ell_\text{AdS}^2)^{-1/2}\\
&=\frac{1}{4}\mathbb W_{j}[a_L,a_R]~.
\end{aligned}
\eeq
Here $j$ labels a lowest-weight representation of $\mfsl$ related to the mass of the bulk field via
\beq
j=\frac{1}{2}\left(1+\sqrt{\msf^2\ell_\text{AdS}^2+1}\right)\equiv\frac{1}{2}\Delta~.
\eeq
 The expression for the Wilson spool, ~\eqref{eq:spooldef}, can be derived for this background. Given \eqref{eq:BTZholo}, we have
\begin{align}\label{eq:BTZspool1}
&\mathbb W_{j}[a_L,a_R]=i\int_{\mc C}\frac{\dd\alpha}{\alpha}\frac{\cos\alpha/2}{\sin\alpha/2}\chi_{j}\left(\frac{\alpha}{2\pi}\hsf_L\right)\chi_{j}\left(-\frac{\alpha}{2\pi}\hsf_R\right)~,
\end{align}
where $\hsf_{L/R}$ are given by \eqref{eq:BTZholo2} and
\beq\label{eq:sl2char}
\chi_j(z)=\mTr_{R_j}\left(e^{i2\pi zL_0}\right)=\frac{e^{i\pi z(2j-1)}}{2\sinh(-i\pi z)}
\eeq
is the character of the lowest-weight representation, $R_j$. The contour, $\mc C$, is given by twice the contour running up the imaginary $\alpha$ axis to the right of zero: $\mc C=2\mc C_+$. This follows from the procedure in \cite{Castro:2023dxp} but assigning an $i\epsilon$ prescription appropriate for representations and holonomies relevant to AdS$_3$.  Further detail of this is also described in the supplementary material.

Because $\tau,\bar\tau\in i\mathbb R$ (equivalently, $\mc L$ and $\bmc L$ are real and positive), all poles in the $\alpha$ integrand in \eqref{eq:BTZspool1} to the right of zero are simple poles at $2\pi\mathbb Z_{>0}$ and arise from the measure, $\frac{\cos\alpha/2}{\sin\alpha/2}$.  We then deform the contour $\mc C$ to the right where the $\alpha$ integrand is damped, picking up the residues of the simple poles. We can then write
\begin{align}\label{eq:AdSspoolans1}
\mathbb W_{j}[a_L,a_R]=&\sum_{n=1}^\infty\frac{e^{-i\pi n(\tau^{-1}-\bar\tau^{-1})(2j-1)}}{n\,\sin\left(\frac{\pi n}{\tau}\right)\sin\left(\frac{\pi n}{\bar\tau}\right)}~.
\end{align}
This can be easily rewritten into
\beq\label{eq:BTZscalarfinal}
\frac{1}{4}\mathbb W_j[a_L,a_R]=\log\prod_{l,\bar l=0}^\infty\left(1-q^{\frac{\Delta}{2}+l}\bar q^{\frac{\Delta}{2}+\bar l}\right)^{-1}~,
\eeq
where $q=e^{-i\frac{2\pi}{\tau}}$ and $\bar q=e^{i\frac{2\pi}{\bar\tau}}$. This matches exactly $\log Z_{\scalar}[\text{BTZ}]$ for a real massive field propagating on the rotating BTZ background \cite{Giombi:2008vd}.

This is the tree-level ($G_N\to0$) contribution to $\gev{\log Z_{\scalar}}$. Promoting the background connections to dynamical fields, $a_{L/R}\rightarrow A_{L/R}$, the Chern-Simons path-integral provides a way forward, in principle, for calculating perturbative $G_N$ corrections to \eqref{eq:BTZscalarfinal}.  The non-compact gauge group and the non-compact background topology make this program still difficult.  However, one may still make progress using large-$k$ Chern-Simons perturbation theory similar to \cite{Fitzpatrick:2016mtp,Besken:2017fsj,Hikida:2017ehf}.  Below we will see that in the context of positive cosmological constant, our ability to calculate $G_N$ corrections is under even better control.

\section{Wilson spools in \texorpdfstring{\lowercase{d}S$_3$}{dS3} gravity} \label{sec:ds}

Our next example will be Chern-Simons gravity with positive cosmological constant, $\Lambda = \ell_{\rm dS}^{-2}$, which describes quantum gravity on  de Sitter space.
We will be working with a Euclidean action given by two copies of the $SU(2)$ Chern-Simons action; its relation to dS$_3$ gravity is via 
\beq\label{eq:SL+SR}
I_\text{EH}-i\delta I_\text{GCS}=-ik_LS_{\rm CS}[A_L]-ik_RS_{\rm CS}[A_R]~,
\eeq
where $I_\text{EH}$ is the Euclidean Einstein-Hilbert action and $I_\text{GCS}$ is the gravitational Chern-Simons action. The Chern-Simons couplings $k_{L,R}$ are now in general complex,
\beq\label{eq:kLRdef}
k_L=\delta+is~,\qquad k_R=\delta-is~,
\eeq
with an imaginary part that is related to Newton's constant, $s=\frac{\ell_\text{dS}}{4G_N}$, and a real part, $\delta$, giving the coefficient of a gravitational Chern-Simons action.  Quantum effects lead to a renormalization in the coupling constants \cite{Witten:1989ip}
\beq\label{eq:kLRrenorm}
k_L\rightarrow r_L=k_L+2~,\qquad k_R\rightarrow r_R=k_R+2~,
\eeq
which amounts to a renormalization of $\delta\rightarrow\hat\delta=\delta+2$.

The above matching is facilitated by relating the gauge fields to the vielbein and spin connection via
\begin{align}\label{eq:ALAR}
A_L=i(\omega^a+\frac{e^a}{\ell_\text{dS}})L_a~,~~ A_R=i(\omega^a-\frac{e^a}{\ell_\text{dS}})\bar L_a~,
\end{align}
where now $\{L_a\}$ and $\{\bar{L}_a\}$ generate the two independent $\mathfrak{su}(2)$'s. We will focus presently on the round $S^3$ saddle whose classical background geometry is 
generated by background connections
\begin{align}\label{eq:S3bgconns}
    a_L=&iL_1\,\dd\rho+i\left(\sin\rho L_2-\cos\rho L_3\right)(\dd\varphi-\dd\tau)~,\nonumber\\
    a_R=&-i\bar L_1\,\dd\rho-i\left(\sin\rho\bar L_2+\cos\rho\bar L_3\right)(\dd\varphi+\dd\tau)~.
\end{align}
These background connections have a singularity at the causal horizon ($\rho=\pi/2$) and the holonomy around that singularity is given by\footnote{The $S^3$ geometry is regular everywhere; the holonomies add up to give a smooth background.}
\bea
\begin{aligned}\label{eq:holalar}
\mathcal P\exp\oint_{\gamma}a_{L/R}&=u_{L/R}^{-1}e^{i2\pi L_3\ms h_{L/R}}u_{L/R}~,
\end{aligned}
\eea
with $u_{L/R}$ periodic group elements and
\beq\label{eq:holalar-v}
\ms h_L=1~,\qquad\ms h_R=-1~.
\eeq

Now let us add matter and test our proposal. We will begin with the tree-level check which evaluates the spool on the background connections:
\begin{align}
\log Z_{\rm scalar}[S^3]&=\frac{1}{4}\mathbb W_j[a_L,a_R]~.
\end{align}
The representation, $R_j$, that appears in $\mathbb W_j$ is a highest-weight representation of $\mfsu$ related to the mass of the scalar field as
\beq\label{eq:nonstandHW}
j=-\frac{1}{2}\left(1+\sqrt{1-\msf^2\ell_\text{dS}^2}\right)~.
\eeq
Note that $j$ is a continuous parameter and can even become complex for large enough masses.  Thus these representations do not correspond to the standard finite dimensional representations of $SU(2)$. Instead they correspond to infinite dimensional ``non-standard" representations whose weight spaces line up with the dS$_3$ quasi-normal mode spectrum \cite{Castro:2020smu,Castro:2023dxp}.  Despite not lying in the standard $SU(2)$ representation theory, the representations obeying \eqref{eq:nonstandHW} can be equipped with an inner product such that all states have positive norm \cite{Castro:2020smu,Castro:2023dxp}.  Additionally they admit well-defined characters:
\be\label{eq:Zukchars} 
\chi_j(z) = \mbox{Tr}_{R_j}\left(e^{2\pi i z L_3}\right)= \frac{e^{i\pi z(2j+1)}}{2 i \sin{(\pi z)}}~.
\ee
We can then express \eqref{eq:dSspool1} in terms of the holonomies \eqref{eq:holalar-v} as
\begin{align}\label{eq:dSspool1}
\mathbb W_{j}[a_L,a_R]=&i\int_{\mc C}\frac{\dd\alpha}{\alpha}\frac{\cos\alpha/2}{\sin\alpha/2}\chi_{j}\left(\frac{\alpha}{2\pi}\hsf_L\right)\chi_{j}\left(-\frac{\alpha}{2\pi}\hsf_R\right)\nonumber\\
=&-\frac{i}{4}\int_{\mc C}\frac{\dd\alpha}{\alpha}\frac{\cos\alpha/2}{\sin^3\alpha/2}e^{i(2j+1)\alpha}~.
\end{align}
Deviating slightly from the procedure for the BTZ computation, the $i\epsilon$ prescription appropriate for de Sitter results in a contour, $\mc C=\mc C_-\cup\mc C_+$, that is, the union of contours running up the imaginary $\alpha$ axis both to the left and right of zero \cite{Castro:2023dxp}. We pull both $\mc C_\pm$ towards the positive real axis to pick up residue of the quadruple pole at $\alpha=0$ as well as twice the residues of the poles at $\alpha\in 2\pi \mathbb Z_{>0}$ along the positive real line; these latter poles are now third-order.  Doing so we find
\begin{align}\label{eq:S3classspool}
\frac{1}{4}\mathbb W_j[a_L,a_R]=&i\frac{\pi(2j+1)^3}{6}-\frac{1}{4\pi^2}\Li3\left(e^{i2\pi(2j+1)}\right)\nonumber\\
&+i\frac{(2j+1)}{2\pi}\Li2\left(e^{i2\pi(2j+1)}\right)\nonumber\\
&-\frac{(2j+1)^2}{2}\Li1\left(e^{i2\pi(2j+1)}\right)~,
\end{align}
where $\Li q(x)=\sum_{n=1}^\infty\frac{x^n}{n^q}$ are polylogarithm functions. 
Upon using \eqref{eq:nonstandHW}, this answer matches precisely the finite contribution to the scalar one-loop determinant on $S^3$; see for instance \cite{Anninos:2020hfj}.  This provides a second non-trivial check of the physical relevance of the Wilson spool.

In the context of de Sitter gravity we can actually make substantial progress in discussing the Wilson spool beyond its classical expectation value.  This is because there exists a library of ``exact techniques" for $SU(2)$ Chern-Simons theories on compact topologies.  These techniques reduce the Chern-Simons path-integral, as well as the expectation values of certain operators, to ordinary integrals.  In the present context, care is needed to alter these methods to accommodate the features necessary for Chern-Simons gravity (i.e. complex levels, non-zero background connections, and non-standard representations). However several exact methods remain amenable to these alterations \cite{Castro:2023dxp}.  For instance, a suitable alteration of Abelianisation~\cite{Blau:1993tv, Blau:2006gh, Blau:2013oha} reduces the expectation value of $\mathbb W_j[A_L,A_R]$ with dynamical connections to a pair of simple integrals \cite{Castro:2023dxp}:
\begin{align}\label{eq:dSWev}
&\frac{1}{4}\gev{\mathbb{W}_j}\nonumber=\frac{i}{4}\,e^{ir_LS_{\rm CS}[a_L]+ir_RS_{\rm CS}[a_R]}\\
&\times\int \dd\sigma_L\dd\sigma_R\,\Big\{e^{i\frac{\pi}{2}r_L\sigma_L^2+i\frac{\pi}{2}r_R\sigma_R^2}\sin^2(\pi\sigma_L)\sin^2(\pi\sigma_R) \nonumber\\
&\times\int_{\mc C}\frac{\dd\alpha}{\alpha}\frac{\cos\alpha/2}{\sin\alpha/2}\chi_{j}\left(\frac{\alpha (\sigma_L+\ms h_L)}{2\pi}\right)\chi_{j}\left(-\frac{\alpha (\sigma_R+\ms h_R)}{2\pi}\right)\Big\}~,
\end{align}
which we can equate to the quantum gravity corrected one-loop determinant about the $S^3$ saddle, $\gev{\log Z_{\scalar}[S^3]}$.  The integral appearing in \eqref{eq:dSWev} is difficult to evaluate analytically, however it can systematically be evaluated in a $s^{-1}$ Taylor expansion.  This provides a controlled procedure to computing $G_N$ corrections to $\log Z_{\scalar}[S^3]$.  These corrections are naturally identified with a mass renormalization.  To extract them one simply computes $\gev{\log Z_{\scalar}[S^3]}$, normalized by the gravitational path integral, using \eqref{eq:dSWev}; the resulting renormalized mass to $O(G_N^2)$ is given by\footnote{In \eqref{eq:mass-R} we have assumed that the (renormalized) gravitational Chern-Simons coupling, $\hat\delta$, does not scale with $G_N$.} 
\beq\label{eq:mass-R}
\msf^2_R\ell^2_\text{dS}=\msf^2\ell^2_\text{dS}+\frac{96}{5}\msf^4\ell^4_\text{dS}e^{-2\pi\,\abs{\msf\ell_\text{dS}}}\,\left(\frac{G_N}{\ell_\text{dS}}\right)^2+\ldots~.
\eeq
Above we have kept the leading term in a large mass expansion, $\msf^2\ell_\text{dS}^2\gg1$, however  \eqref{eq:dSWev} provides an expression for the $G_N^2$ renormalization of the mass that can be calculated analytically \cite{Castro:2023dxp}. We emphasize that this is a concrete predictive statement about how dynamical quantum gravity renormalizes quantum field theory.

\section{Discussion}\label{sec:disc}

We have introduced a new object, the Wilson spool, which allows one to have matter fields in the Chern-Simons formulation of three-dimensional quantum gravity while keeping manifest the key topological aspects of gravity. To test this object,  we have shown that, for $G_N\rightarrow 0$, it reproduces correctly one-loop determinants of massive scalar fields on a curved background. The proposal works for spacetimes that are widely different, such as the spinning BTZ black hole and Euclidean de Sitter spacetime. 

If our proposal merely provided a way to match onto known one-loop determinants, it would have limited utility. However, it is also possible to use this techniques to make a \emph{prediction} for quantum corrections to $\log Z_{\rm grav}$, having allowed for quantum fluctuations of the metric around a classical background. For certain AdS geometries (see, e.g.,~\cite{Fitzpatrick:2016mtp}), the $1/c$ corrections to Wilson lines have been computed in a holographic setting. Our techniques give us a way to extend this beyond the holographic setting to quantum gravity in spacetimes like de Sitter, which are more realistic approximations of our observed universe but where techniques of holography are largely out of reach.

There are two future directions in this area that we would like to highlight. A more in depth discussion is presented in \cite{Castro:2023dxp}.

\paragraph{Massive spinning fields.}
We have focused on massive scalar fields in this letter, for which we have a direct construction of the spool.  But both AdS$_3$ and dS$_3$ admit massive fields of arbitrary integer spin. We can ask if the Wilson spool is of utility in coupling these excitations to dynamical gravity. We do not presently have a first-principles construction for $\mathbb W$ applied to spinning fields (as we do for scalars).  However, in the context of AdS$_3$, the definition of the Wilson spool, \eqref{eq:spooldef}, can be intuitively extended to reproduce the one-loop determinant for a massive spinning field on the rotating BTZ background. The details of this can be found in the supplementary material. This is a strong indication that $\mathbb W$ is useful for the physics of spinning fields. We can also make a first pass application of $\mathbb W$ to spinning fields in de Sitter; this relies on further guess work since the appropriate non-standard $\mfsu$ representations have not yet been identified. In this case, while we can reliably reproduce certain contributions to the one-loop determinant, $\mathbb W$ currently misses an important ``edge subtraction" arising from properly subtracting normalizable transverse-traceless zero modes on $S^3$ \cite{Anninos:2020hfj}. A proper derivation of the spool to capture the physics of spinning fields will be addressed in future work.

\paragraph{Sum over topologies.} Here we have kept the topology of the background, and holonomies of the connections, fixed. It is of great interest to allow for these non-perturbative contributions in \eqref{eq:Zmatter}, albeit it comes with difficulties: if only metric degrees of freedom are incorporated, both the AdS$_3$ thermal partition function \cite{Maloney:2007ud,Keller:2014xba} and the Euclidean dS$_3$ path integral \cite{Castro:2011xb} suffer from pathologies. But it is also expected that adding matter can fix some of these problems \cite{Benjamin:2020mfz}. It would interesting to  investigate the fate of the Wilson spool under these pathologies, i.e., loop in matter in the sum over manifolds and quantify its imprint. This is specially of interest in dS$_3$, where a holographic dictionary is still nascent, but the Wilson spool gives a concise path to quantify the effect of fields at all orders in $G_N$.

\begin{acknowledgments}
\section*{Acknowledgments}
We would like to thank Kurt Hinterbichler and Sergei Cherkis for useful comments and discussions. JRF thanks the University of Amsterdam for hospitality. The work of AC and JRF has been partially supported by STFC consolidated grant ST/T000694/1. The work JRF has been also partially supported by Simons Foundation Award number 620869.  IC has been partially supported by the ERC starting grant H2020 ERC StG No.640159. CZ has been supported by a UM Duluth Higholt Professorship.

\end{acknowledgments}

\vspace{10 pt}


%

\appendix{}

\titlepage

\pagebreak

\setcounter{page}{1}

\begin{center}
{\LARGE Supplemental Material}
\end{center}

\section{A. Details on constructing the Wilson spool in AdS$_3$} 
The construction of $\mathbb W_j$ as described in \cite{Castro:2023dxp} makes use of the physical reasoning established by Denef, Hartnoll, and Sachdev (DHS) \cite{Denef:2009kn}. Namely, the one-loop determinant
\beq
\left(Z_\text{scalar}\right)^2=\left(\text{det}\left(-\nabla^2+\msf^2\right)\right)^{-1}~,
\eeq
interpreted as a meromorphic function of the mass, is equal to the rational function containing the same zeros and poles (up to multiplication by an entire holomorphic function).  $\left(Z_\text{scalar}\right)^2$ will have poles in the complex $\msf^2$ plane on solutions to
\beq
\left(-\nabla^2+\msf^2\right)\phi=0~,
\eeq
subject to appropriate boundary conditions.

For asymptotically AdS backgrounds, we interpret this statement in the language of $\mfsl$ representation theory by noting that there exist vector fields, $\{\zeta_0,\zeta_+,\zeta_-\}$ and $\{\bar\zeta_0,\bar\zeta_+,\bar\zeta_-\}$, spanning $\mfsl_L$ and $\mfsl_R$, respectively, whose summed Casimirs act as the Laplacian \cite{Castro:2018srf}:
\beq
\ell_\text{AdS}^2\nabla^2=\zeta_+\zeta_-+\zeta_-\zeta_+-2\zeta_0^2+\bar\zeta_+\bar\zeta_-+\bar\zeta_-\bar\zeta_+-2\bar\zeta_0^2=2c_{2,L}+2c_{2,R}~.
\eeq
The DHS procedure instructs us to write
\beq
\left(Z_\text{scalar}\right)^2=\left(\text{det}(-2c_{2,L}-2c_{2,R}+\msf^2\ell^2_\text{AdS})\right)^{-1}
\eeq
as the simple product function containing the same poles.  These poles lie along weights of scalar representations of $\mfsl_L\oplus\mfsl_R$ satisfying $c_{2,L}=c_{2,R}=\frac{1}{4}\msf^2\ell^2_\text{AdS}$.  These are highest/lowest-weight representations with $j_L=j_R\equiv j$ and
\beq\label{eq:jroots}
j=\frac{1}{2}\left(1\pm\sqrt{\msf^2\ell^2_\text{AdS}+1}\right)~,
\eeq
where $j$ is a lowest (highest)-weight if it is positive (negative). However, unlike in de Sitter, it is not the case that both roots of \eqref{eq:jroots} can contribution poles to $\left(Z_\text{scalar}\right)^2$.  This is because we must impose boundary conditions: in particular, at asymptotic infinity we fix a Dirichlet boundary condition on the scalar field, $\Phi$, with normalizable fall-off. This then fixes $\Phi$ to correspond to a unique lowest-weight representation.  For the rest of this letter we will fix this to the lowest-weight representation with the positive root, $j=\frac{1}{2}\left(1+\sqrt{\msf^2\ell^2_\text{AdS}+1}\right)$.  We should note that for $-1<\msf^2\ell^2_\text{AdS}<0$ the other solution is also positive and so is also compatible with a normalizable boundary condition; the results for this ``alternative quantization" follow exactly the same procedure laid out below.

Additionally we must impose periodicity along the non-contractible spatial cycle, $\gamma_\varphi$.  For a field living in the $R_j\otimes R_j$ representation of $\mfsl_L\oplus\mfsl_R$, its parallel transport along this cycle is
\beq
\phi\rightarrow R_j\left(\mc Pe^{\oint_{\gamma_\varphi} a_L}\right)\cdot\phi\cdot R_j\left(\mc Pe^{-\oint_{\gamma_\varphi} a_R}\right)~,
\eeq
and so a weight $(\lambda_L,\lambda_R)\in R_j\otimes R_j$ can contribute a pole to $\left(Z_\text{scalar}\right)^2$ if
\beq
\lambda_L\hsf_L-\lambda_R\hsf_R=\pm n~,\qquad n\in\mathbb Z~,
\eeq
where $\hsf_{L/R}$ are the holonomies of the background connections, \eqref{eq:BTZholo2}.  The $\pm$ arises from imposing periodicity on both the left and right-propagating modes.  Alternatively, it arises from choosing the sign of the square-root in arriving at \eqref{eq:BTZholo2}.  In the metric language this is the same as imposing periodicity on both ingoing and outgoing quasi-normal modes \cite{Denef:2009kn,Castro:2017mfj}.  Note that in de Sitter this contribution is accounted for by including the shadow representation \cite{Denef:2009kn}.
Thus we arrive at the following expression for $\left(Z_\text{scalar}\right)^2$:
\begin{align}
\left(Z_\text{scalar}\right)^2=\prod_{\overset{(\lambda_L,\lambda_R)}{\in R_j\otimes R_j}}\prod_{n\in\mathbb Z}\left(|n|-\lambda_L\hsf_L+\lambda_R\hsf_R\right)^{-1}\left(|n|+\lambda_L\hsf_L-\lambda_R\hsf_R\right)^{-1}~.
\end{align}
We now take the log.  We will implement this with a Schwinger parameter $\log M=-\int_\times^\infty \frac{\dd\alpha}{\alpha}e^{-\alpha M}$, where in the lower limit we indicate a need to regulate the UV divergence at $\alpha\sim 0$.  Additionally we will need to regulate the sum over representation weights $(\lambda_L,\lambda_R)$.  Both of these can be taken care of simultaneously through an $i\epsilon$ prescription which we implement now.  Noting $\sum\limits_{n\in\mathbb Z}e^{-|n|\alpha}=\frac{\cosh\alpha/2}{\sinh\alpha/2}$, we write
\beq
\log Z_{\scalar}=\frac{1}{2}\int_\times^\infty \frac{\dd\alpha}{\alpha}\frac{\cosh\alpha/2}{\sinh\alpha/2}\sum_{\lambda_L,\lambda_R}\left(e^{\alpha\left(\lambda_L\hsf_L-\lambda_R\hsf_R\right)}+e^{-\alpha\left(\lambda_L\hsf_L-\lambda_R\hsf_R\right)}\right)~.
\eeq
Note that $\hsf_L$ is equal to $i$ times a positive quantity, while $\hsf_R$ is $i$ times a negative quantity.  Also, since $R_j$ is a lowest-weight representation of $\mfsl$, all $\lambda_{L/R}$ appearing in the sum are real and positive.  It then follows that the first (second) sum converges by giving $\alpha$ a small positive (negative) imaginary part.  
Under the replacement of integration variable, $\alpha\rightarrow-\alpha$, in the second sum, we can write the regulated $\log Z_{\scalar}$ succinctly as
\beq
\log{Z_{\scalar}}=\frac{1}{2}\int_{-\infty+i\epsilon}^{\infty+i\epsilon} \frac{\dd\alpha}{\alpha}\frac{\cosh\alpha/2}{\sinh\alpha/2}\sum_{\lambda_L,\lambda_R}e^{\alpha\left(\lambda_L\hsf_L-\lambda_R\hsf_R\right)}~.
\eeq
We recognize the sum over representation weights times holonomies as trace path-ordered exponentials of the background connections. Finally, we make the integration variable replacement $\alpha\rightarrow i\alpha$ to write
\beq
\log Z_{\scalar}=\frac{i}{4}\int_{2\mc C_+}\frac{\dd\alpha}{\alpha}\frac{\cos\alpha/2}{\sin\alpha/2}\mTr_{R_j}\left(\mc Pe^{\frac{\alpha}{2\pi}\oint_{\gamma_\varphi}a_L}\right)\mTr_{R_j}\left(\mc Pe^{-\frac{\alpha}{2\pi}\oint_{\gamma_\varphi}a_R}\right)\equiv \frac{1}{4}\mathbb W_j[a_L,a_R]~.
\eeq
Thus appears the Wilson spool. The contour $\mc C_+$ runs upwards along the imaginary $\alpha$ axis to the right of zero.  We write it conventionally with the factor of two to uniformize the notation with the de Sitter spool (where the ingoing vs. outgoing periodicity constraints give rise to distinct contours \cite{Castro:2023dxp}). We derived $\mc C_+$ by ensuring representation sums converge, however, in the body of the letter we verify {\it a posteriori} that it also regulates the UV divergence in $\log Z_{\scalar}$.

\section{B. Massive spinning fields}

We provide additional details on a ``first-pass" attempt to incorporate spinning fields into the Wilson spool.  This is easiest to address in the context of AdS$_3$, where massive particles of spin $\ssf$ are already included in $\mfsl_L\oplus\mfsl_R$ representation theory through lowest-weights
\beq
j^+=\frac{\Delta+\ssf}{2}~,\qquad j^-=\frac{\Delta-\ssf}{2}~.
\eeq
Noting that in \cite{Castro:2023dxp} that the construction of $\mathbb W$ arises from a sum of Casimirs, $c_{2,L}+c_{2,R}$, it is natural to conjecture that both contributions, $(j_L,j_R)=(j^+,j^-)$ and $(j_L,j_R)=(j^-,j^+)$, should be included the spool since both choices have the same sum of Casimirs:
\begin{align}\label{eq:spinspoolguess}
\mathbb W_{j^+,j^-}[A_L,A_R]=i\int_{\mc C}\frac{\dd\alpha}{\alpha}\frac{\cos\alpha/2}{\sin\alpha/2}\sum_{\pm}\mTr_{R_{j^\pm}}\left(\mc P e^{\frac{\alpha}{2\pi}\oint A_L}\right)\mTr_{R_{j^\mp}}\left(\mc Pe^{-\frac{\alpha}{2\pi}\oint A_R}\right)~.
\end{align}
Indeed doing so, we can calculate for the BTZ background, \eqref{eq:BTZbgconns}, that the tree-level contribution to the spinning spool is
\begin{align}\label{eq:BTZspinfinal}
\frac{1}{4}\mathbb W_{j^+,j^-}[a_L,a_R]=&\sum_{\pm}\sum_{n=1}^\infty\frac{1}{n}\frac{q^{\frac{n}{2}(\Delta\pm\ssf)}\bar q^{\frac{n}{2}(\Delta\mp\ssf)}}{(1-q^n)(1-\bar q^n)}=-\sum_{\pm}\sum_{l,\bar l=0}^\infty\log\left(1-q^{\frac{1}{2}(\Delta\pm\ssf)+l}\bar q^{\frac{1}{2}(\Delta\mp\ssf)+\bar l}\right)~,
\end{align}
which is the correct one-loop determinant for a massive spinning field on the rotating BTZ background \cite{David:2009xg,Datta:2011za,Castro:2017mfj}. We stress that, as of present, we do not have a first principles construction for $\mathbb W_{j^+,j^-}$ to justify the above calculation. However in the context of AdS$_3$ we have reason to trust the above: massive spinning fields in AdS$_3$ have only two real physical polarizations \cite{David:2009xg} which we can package into $\mfsl$ lowest-weights $(j^+,j^-)$. If the one-loop determinant continues to depend only on the summed Casimirs, as it does for scalar fields, then is hard to imagine any other appropriately group theoretic object to construct besides \eqref{eq:spinspoolguess}.

Trying to extend this reasoning to de Sitter gravity is more subtle. Part of the issue is that we already need novel $\mfsu$ representations to capture the physics of scalar excitations; as of yet we have not constructed the non-standard representations necessary to accommodate $\ssf\neq0$.  We view this as technical hurdle as opposed to a conceptual no-go. 
As a first pass, we can analytically continue the $\mfsu$ weights to accommodate spin via
\beq
j^+=-\frac{\Delta+\ssf}{2}~,\qquad j^-=-\frac{\Delta-\ssf}{2}~.
\eeq
Investigating \eqref{eq:spinspoolguess} for the classical $S^3$ background, \eqref{eq:S3bgconns}, we can implement this analytic continuation at the level of the characters. We find that $\mathbb W_{j^+,j^-}[a_L,a_R]$ is, in fact, $\ssf$ independent and essentially doubles the scalar answer, \eqref{eq:S3classspool}. This almost reproduces one piece of the one-loop determinant of massive spinning fields on $S^3$,
\beq\label{eq:dSspinmismatch}
\log Z_{\Delta,\ssf}[S^3]=\frac{1}{4}\mathbb W_{j^+,j^-}[a_L,a_R]-\log Z_\text{edge}~,
\eeq
however it misses an additional ``edge" subtraction
\beq
\log Z_\text{edge}=\ssf^2\log\left(\frac{e^{i\pi(\Delta-1)}}{1-e^{i2\pi(\Delta-1)}}\right)~,
\eeq
which arises from properly subtracting normalizable transverse-traceless zero modes \cite{Anninos:2020hfj}. We hope that a suitable modification of $\mathbb W$ can capture this edge term, which we leave for future work.

\end{document}